\documentclass[sigconf,edbt]{acmart-edbt2018}
\usepackage[utf8]{inputenc}

\usepackage{booktabs} 

\usepackage{scrhack}

    \usepackage{listings}

    \usepackage{graphicx}  
    \usepackage{subfig}  
    \usepackage{amsmath}  
    \usepackage{amsthm}   
    \usepackage{amsfonts} 
    \usepackage{calc}  

    \usepackage{algorithm,algpseudocode}
    \usepackage{bm}  
    \usepackage{tikz}
    \usetikzlibrary{positioning}  

    \renewcommand{\eqref}[1]{Equation~(\ref{#1})}

    \definecolor{darkgreen}{rgb}{0.0, 0.5, 0.0}
    \definecolor{UniBlue}{RGB}{0, 74, 153}
    \definecolor{UniRed}{RGB}{193, 0, 42}
    \definecolor{UniGrey}{RGB}{154, 155, 156}

    \tikzset{>=latex}  

    \algnewcommand\algorithmicforeach{\textbf{foreach}}
    \algdef{S}[FOR]{ForEach}[1]{\algorithmicforeach\ #1\ \algorithmicdo}
\usepackage{amsthm}

\setcopyright{rightsretained}

\acmDOI{}

\acmISBN{978-3-89318-078-3}

\acmConference[EDBT 2018]{21st International Conference on Extending Database Technology (EDBT)}{March 26-29, 2018}{Vienna, Austria} 
\acmYear{2018}

\begin{document}

\title{PRoST: Distributed Execution of SPARQL Queries Using Mixed Partitioning Strategies}


\author{Matteo Cossu}
\orcid{}
\affiliation{%
  \institution{Institute for Computer Science\\University of Freiburg}
  \postcode{79110}
}
\email{elcossu@gmail.com}

\author{Michael Färber}
\orcid{}
\affiliation{%
  \institution{Institute for Computer Science\\University of Freiburg}
  \postcode{79110}
}
\email{michael.faerber@cs.uni-freiburg.de}

\author{Georg Lausen}
\orcid{}
\affiliation{%
  \institution{Institute for Computer Science\\University of Freiburg}
  \postcode{79110}
}
\email{lausen@informatik.uni-freiburg.de}

\renewcommand{\shortauthors}{M. Cossu et al.}

\begin{abstract}
The rapidly growing size of RDF graphs in recent years necessitates distributed storage and parallel processing strategies. 
To obtain efficient query processing using computer clusters a wide variety of different approaches have been proposed. 
Related to the approach presented in the current paper are systems built on top of Hadoop HDFS, for example using Apache Accumulo or using Apache Spark. We present a new RDF store called PRoST (Partitioned RDF on Spark Tables) based on Apache Spark. PRoST introduces an innovative strategy that combines the Vertical Partitioning approach with the Property Table, two preexisting models for storing RDF datasets. We demonstrate that our proposal outperforms state-of-the-art systems w.r.t. the runtime for a wide range of query types and without any extensive precomputing phase.
\end{abstract}
%
%


\maketitle

    \section{Introduction}
Organizations and institutions increasingly see a need to represent their data in a semantically-structured way~\cite{FaerberSWJ2016} and, thus, rely on the RDF data model. As the data represented in RDF constantly grows in size, storing and querying these very large RDF graphs becomes a major challenge.
%
In order to increase the query efficiency, many different approaches have been proposed e.g. \cite{abadi2007scalable, lee2013scaling, punnoose2012rya, schatzle2016s2rdf}.
Most of the current solutions 
use DBMS-based systems and map SPARQL queries to SQL queries for the retrieval. 
Besides the systems which are implemented for single-node machines, such as RDF-3X~\cite{neumann2008rdf}, Sesame~\cite{broekstra2002sesame} and Jena~\cite{wilkinson2006jena}, 
systems designed for distributed environments are increasingly being used.
They allow to scale up the storage and to provide parallel query execution capabilities, which is essential to handle very large data. 
Systems of distributed environments use in particular Hadoop technologies: S2RDF~\cite{schatzle2016s2rdf} and SPARQLGX~\cite{graux2016sparqlgx} are implemented on top of Apache Spark, Rya~\cite{punnoose2012rya} on top of Apache Accumulo, and Sempala~\cite{schatzle2014sempala} on top of Impala.

Typically, these systems are optimized for specific query patterns and their data loading times are often sacrificed for better querying performance.
Thus, there is need for a distributed RDF store with better performance on a wide range of query types, without renouncing a rapid loading phase.

In this work, we describe the approach of PRoST\footnote{The source code of PRoST is online available at  \url{https://github.com/tf-dbis-uni-freiburg/PRoST}.} (Partitioned RDF on Spark Tables), that aims to improve the efficiency of queries on RDF data, thereby covering a wide range of query types.
Instead of building a standalone system, such as \cite{PotterMNH16}, AdPart~\cite{abdelaziz2017combining} or TriAD~\cite{gurajada2014triad}, PRoST is based on reliable Hadoop technologies for storing and processing the data and therefore its performance depends more on the general approach than on the details of the implementation (e.g. memory management, network protocols).
In this paper, we focus on querying RDF efficiently by means of Apache Spark\footnote{\url{https://spark.apache.org}.}, as it emerged as an important component of the Hadoop ecosystem: It offers general purpose data processing functions that exploit efficiently the main memory --- differently from MapReduce that relies mainly on disk operations. This characteristic makes Spark very fast in practice and able to compute complex queries on large RDF graphs.
In particular, it is possible to obtain high querying performances using Spark SQL, a module of Spark that adds well-established DBMS techniques to the distributed computation framework.
 
Our main contribution is proposing \textit{a new storage model for loading RDF graphs into Hadoop}, 
including an appropriate strategy for translating SPARQL queries into Spark execution plans. 
We show that combining the so far separately applied strategies for representing RDF data in Hadoop --- Vertical Partitioning \cite{abadi2007scalable} and the strategy of using Property Tables \cite{wilkinson2006jena} --- can be implemented with relatively little effort 
and 
leads to superior querying performances in many cases, as our evaluation results show.  

The paper is structured as follows:
In Section~\ref{sec:relatedwork}, we present an overview of similar systems for querying RDF.
In Section~\ref{sec:approach}, we outline our approach for modeling the data, our strategy for translating SPARQL queries, and the techniques for improving the performances of PRoST.
We summarize our evaluation results in Section~\ref{sec:experiments} 
and, in Section~\ref{sec:conclusionfuturework}, we conclude and present ideas for future implementations of PRoST.

    \section{Related work}
\label{sec:relatedwork}
Due to space limitations, in the following we restrict ourselves to Hadoop-based RDF querying approaches, as they are the most similar to our system PRoST. 
\textit{S2RDF}~\cite{schatzle2016s2rdf} is a "SPARQL processor" based on Spark SQL. It introduces a data partitioning approach that extends Vertical Partitioning~\cite{abadi2007scalable} with additional tables, containing precomputed semi-joins.
As a consequence, many intermediate results of queries are already computed and can be used to shorten the retrieval time. 
However, S2RDF trades off the performances with disk space and loading time. For datasets with a large number of properties (e.g., DBpedia \cite{auer2007dbpedia}), the time required may make the loading unfeasible.

\textit{SPARQLGX}~\cite{graux2016sparqlgx} is another system for distributed SPARQL queries that uses Vertical Partitioning. SPARQLGX compiles the queries directly into Spark operations. The system relies entirely on its own statistics to optimize the computation, in particular for the order of the joins. Differently from S2RDF and PRoST, SPARQLGX does not use Spark SQL.

\textit{Rya}~\cite{punnoose2012rya} is a popular RDF management system, developed by an active open-source community and currently an incubating project at Apache.
It is built on top of Apache Accumulo\footnote{\url{https://accumulo.apache.org/}.}, a distributed key-value datastore for Hadoop. Since Accumulo keeps all its information sorted and indexed by key, Rya stores whole RDF triples as keys. For this reason, single triples or short ranges can be accessed very fast.
However, considering that the triple table model of Rya contains three columns, it requires to replicate the data multiple times in order to exploit the indexes over all the possible elements.
Rya uses some query optimization, e.g. joins reordering, but it lacks of the powerful in-memory data processing that make, in practice, other systems faster. 
    \section{Approach}
\label{sec:approach}
The main idea behind PRoST is to store the data twice, partitioned in two different ways. Each one of the two data structures can be more beneficial (in terms of performance) than the other regarding certain query types, mainly because of the different location of the data in the cluster. Our system tries to exploit this characteristic. 
We do not limit ourselves to choose the single best model out of these two for each query, but we split a query in several parts, such that each sub-query can be executed with the most appropriate approach. At the end, all the parts can be joined together to produce the final result. In this way, our approach achieves better performance than when running only on one of the two representations, as it leverages the synergy effects of both models.

\subsection{PRoST Data Model}
\label{datamodels}
The two data structures we use to store an RDF graph are \textit{Vertical Partitioning} and \textit{Property Table}. 
\textit{Vertical Partitioning} (VP) is an approach proposed by Abadi et al.~\cite{abadi2007scalable}. It has a strong popularity among RDF storage systems and it is the main choice of the state-of-the-art approaches S2RDF~\cite{schatzle2016s2rdf} and SPARQLGX~\cite{graux2016sparqlgx}.
The main concept is to create a table for each distinct predicate of the input graph, containing all the tuples (subject, object) that are connected by that predicate. 
The vertical partitioning approach is at the same time powerful and simple, but it has the drawback that it needs a number of joins proportional to the number of triple patterns in the query (precisely, the number of triple patterns minus one). Even with this disadvantage, because the VP tables are narrow and often small, VP is a valid and practical choice.

The \textit{Property Table} (PT) is a data scheme that was introduced in the implementation of Jena2~\cite{wilkinson2006jena}, an influential toolkit for RDF graph processing, and Sempala~\cite{schatzle2014sempala}, another SPARQL processor built on top of the Impala framework. 
A PT consists of a unique table where each row contains a distinct subject and all the object values for that subject, stored in columns identified by the property to which they belong.

The most important advantage of PT, when compared with VP, is that several joins can be avoided when some of the triple patterns in a query share the same subject. In this case, part of the query can be executed by a simpler \texttt{select} operator. Therefore, this approach is beneficial for queries where all the triple patterns have the same subject variable, usually called \textit{star} queries.
However, as observed by Abadi et al.~\cite{abadi2007scalable}, the PT has some issues, in particular the number of NULLs and the multi-valued attributes. Since not every possible pair subject-predicate has a corresponding object value, the PT potentially contains a very large number of NULLs. We solve this problem by storing the table in \textit{Parquet}~\footnote{http://parquet.apache.org}, a format that uses run-length encoding. 
The other problem of the PT is the presence of multi-valued properties, i.e., when more than one different object value exists for at least one subject. In this case, the values are stored using lists that need to be flattened when executing operations on these attributes. This particular situation produces an overhead in respect to the VP model, significant if we use the PT for a single attribute, but negligible if compared to the benefit of avoiding more than one join to compute the same query.

We have several advantages (e.g. efficient compression) in using the columnar format \textit{Parquet} to store the PT, but we then introduce also the undesirable possibility that the data belonging to a single subject could be split into different nodes. In order to keep benefiting from the structure of the property table, we partition horizontally on the subject column. In this way, we ensure that every row is stored entirely in the same node.

\subsection{Query Strategies}
\label{querystrategies}
Having two data structures at the same time requires an abstraction to represent the queries. For this reason, we define an intermediate format that we call \textit{Join Tree}. Each node of this tree represents a sub-query extracted from a VP table or from the PT. Using Spark, PRoST can execute a query by calculating the intermediate results from the nodes and then joining them together in a bottom-up fashion.

\begin{figure}[tb]
    \centering
    \includegraphics[width=0.63\linewidth]{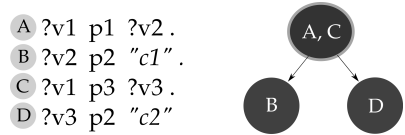}
    \vspace{-0.5em}
    \caption{Example of a conjunction of triple patterns translated to a Join Tree. The root node uses the Property Table, the others Vertical Partitioning.}
    \label{fig:joinTree_PT}
\end{figure}

We now explain how we translate SPARQL queries into the \textit{Join Tree} format.
For the sake of clarity, we consider queries with a unique \textit{basic graph pattern} without filter, which are a conjunction of triple patterns.
The triple patterns with the same subject are grouped together and translated into a single node with a special label, denoting that we should use the property table in this case. All the other groups with a single triple pattern are translated to nodes that will use the vertical partitioning tables. An example is shown in Figure~\ref{fig:joinTree_PT}.
There could be a join edge between every pair of nodes that shares a variable, it means that more than one  \textit{Join Tree} translation is possible for a single query. Since the tree structure influences the final performances, choosing carefully the \textit{Join Tree} is important for the quality of the system. For this reason, we choose a tree guided by simple statistics, as explained in the following section.

\subsection{Statistics-based Optimization}
\label{sec:optimization}
In the context of our work, joins are the most expensive operators because they need large portion of the data to be shuffled across the network.
The order of the joins is important to limit this issue, and therefore to speed up the calculation. An effective way is sorting the joins using statistics of the input graph.
The \textit{Join Tree} of a query decides indirectly the order in which the operations are computed. For example, the leaves are computed first, and the root node is the final one.
The statistics we use, simple but effective in practice, are (1) the total number of triples and (2) the number of distinct subjects for each predicate. 
They are calculated during the loading phase without any significant overhead.

As a general rule of thumb, we want to compute the joins involving few data or with a high selectivity first (high priority), and finally the ones with many tuples (low priority).
The priority value of a node of the Join Tree is decided by the following criteria:
\begin{itemize}
  \item Triple patterns containing literals are scored with the highest priority. The presence of a literal is a strong constraint that limit the number of resulting tuples. Therefore, it is a good approach to push down these kind of nodes.
  \item A triple pattern of which the underlying data contains many tuples will be scored proportionally. For instance, a triple with the highest number of tuples will have also the lowest priority and it will be the root of the tree. This number is also adjusted according the number of distinct subjects for that predicate.
  \item The priority of a node containing data belonging to the Property Table is scored taking in account all its triple patterns. However, the presence of a triple pattern with a literal is weighted heavily.
\end{itemize}
In addition to our effort, Spark SQL's Catalyst Optimizer~\cite{armbrust2015spark} uses its internal heuristics to improve query performances further. The trees are not substantially changed, but Spark intervenes in producing optimized physical plans, since it knows the concrete location of the data on the cluster. 
In particular, the optimizer can choose the type of joins to perform, for example if one of the relations involved is small, a broadcast join will be performed.
    \section{Evaluation Results}
\label{sec:experiments}

\subsection{Benchmark Environment}
We perform our tests on a small cluster of 10 machines connected via Gigabit Ethernet connection. Each machine is equipped with 32GB of memory, 4TB of disk space and with a 6 Core Intel Xeon E5-2420 processor. 
The cluster runs Cloudera CDH 5.11.0 on Ubuntu 14.04 and Spark 2.1.0. Since one machine is the master, Spark uses only 9 workers. The executor's memory is 21GB.
We use the WatDiv~\cite{alucc2014diversified} dataset provided by the Waterloo SPARQL Diversity Test Suite 7. It is developed in order to measure how an RDF data management system performs across a wide variety of SPARQL queries with varying characteristics and has already been applied in the evaluations of the comparable systems.
Given WatDiv we generate a dataset containing around 100 Million RDF triples (5.16 GB) and we evaluate PRoST with the given \textit{WatDiv basic query set}. It contains queries of varying shape and selectivity in order to model different scenarios. The queries are grouped into the following subsets:
\begin{itemize}
    \item \textbf{C}: Complex shaped queries.
    \item \textbf{F (F1, F2, F3, F4, F5)}: Snowflake shaped queries.
    \item \textbf{L (L1, L2, L3, L4, L5)}: Linear shaped queries.
    \item \textbf{S (S1, S2, S3, S4, S5, S6, S7)}: Star shaped queries. 
\end{itemize}

\subsection{Loading}
As we can see from Table~\ref{tab:loadingOtherSystems}, the storage size used by PRoST decreases when being stored in HDFS. This is a direct consequence of the compression used by the \textit{Parquet} format.

\begin{table}[tb]
\begin{center}
    \begin{tabular}{|l|r|r|r|}
        \hline
        \textbf{System}  & \textbf{Size} & \textbf{Time}\\ \hline \hline
        PRoST & 2.1~GB & 25m 32s\\
        SPARQLGX & 0.9~GB & 20m 01s\\
        S2RDF & 6.2~GB & 3h 11m 44s\\
        Rya & 3.1~GB & 41m 32s\\
        \hline
    \end{tabular}
    \end{center}
    \caption[Table caption]{ Size and loading times using WatDiv100M}
    \label{tab:loadingOtherSystems}
\end{table}

The database size of S2RDF is the largest in this analysis, since its model not only needs the creation of VP tables, but also many additional ones containing the results of its precomputations. SPARQLGX occupies minimal space because it uses only Vertical Partitioning, where instead PRoST requires double SPARQLGX's size for having in addition the Property Table.
The loading time of PRoST is similar to SPARQLGX and around an order of magnitude shorter than S2RDF. Similarly to what we said before for the database size, S2RDF's model requires longer times because of its extensive precomputations.

\subsection{Vertical Partitioning vs Mixed Strategy}
\begin{figure}[tb]
    \centering
    \includegraphics[width=\linewidth]{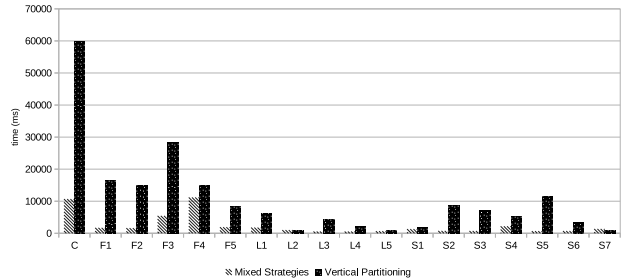}
    \vspace{-1.5em}
    \caption{Querying time for WatDiv100M with only Vertical Partitioning and with the mixed strategy.}
    \label{fig:VPvsPT}
\end{figure}

One of our central points is the introduction of the Property Table alongside Vertical Partitioning. As a consequence, we evaluated the impact of this addition on query performances.\\
In Figure~\ref{fig:VPvsPT}, we show the times to compare the query set from WatDiv100M with Vertical Partitioning only and with the additional support of the Property Table.

\begin{figure*}[tb]
    \centering
    \includegraphics[width=0.68\textwidth]{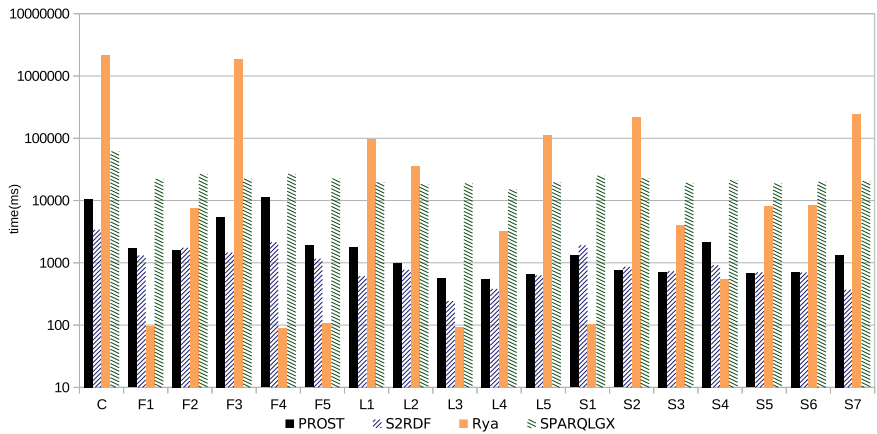}
    \vspace{-0.6em}
    \caption{Querying time for WatDiv100M with PRoST, S2RDF, Rya and SPARQLGX. The scale is logarithmic.}
    \label{fig:resFFA}
\end{figure*}

The chart shows clearly that the introduction of the Property Table has a strong positive impact on performances. For almost every type of query this version outperforms abundantly the simple Vertical Partitioning approach.
The new strategy is effective for Star queries~(S), where most of the triples share the same subject, as well as Complex~(C) and Snowflake~(F) queries, that have more than one different subject.
For some of the Linear queries~(L), the results are very similar between the two versions. This result comes from the fact that Linear queries contain mostly triples with distinct subject variables, translated using mostly Vertical Partitioning.
 
\subsection{PRoST vs Other Systems}
To better evaluate our system we compared it in the same environment with other similar solutions: S2RDF, Rya and SPARQLGX.\\
In Figure~\ref{fig:resFFA} we show the querying times on the WatDiv100M dataset with the other systems and our implementation.
Note that we used a logarithmic scale, since the large differences would prevent us from visualizing these results well.
Compared to S2RDF, PRoST is faster for the queries F2, S1, S3 and S5 but slower for the remaining queries. In particular, for the complex queries (C) and some of the snowflake ones (F3, F4) the S2RDF's approach makes it faster by a considerable margin. These differences can be explained with the extensive
precomputations of S2RDF, that heavily decrease the processing time for joins between VP tables. However we have to keep in mind that S2RDF can reach these results because of a possibly large loading time, that, for some datasets, may make it less attractive in practice. In contrast, PRoST relies on a faster loading phase and its performances does not depend on the particular input graph, i.e. number of predicates.
\begin{table}[tb]
\begin{center}
    \begin{tabular}{|l|r|r|r|r|}
        \hline
        \textbf{Queries}  & \textbf{ProST} & \textbf{S2RDF} & \textbf{Rya} & \textbf{SPARQLGX} \\ \hline \hline
        Complex & 9,364ms & 3,392ms & 2,195,322ms & 61,363ms \\
        Snowflake & 5,923ms & 1,564ms & 369,016ms & 24,046ms \\
        Linear & 2,419ms & 527ms & 49,044ms & 18,254ms \\
        Star & 1,195ms & 884ms & 6,9606ms & 2,1046ms \\
        \hline
    \end{tabular}
    \end{center}
    \caption[Table caption]{Average querying time grouped by type of query.}
    \label{tab:averageQueries}
    \vspace{-5mm}
\end{table}

For some queries, Rya is very fast and outperforms PRoST but in other cases it suffers from significant slowdowns, that are several orders of magnitude slower.
Therefore, its average query time is the worst of the systems considered, as shown in Table~\ref{tab:averageQueries}. 
The queries in which Rya performs very well have in common that their computation involves only few intermediate results. As a matter of fact, Rya uses powerful indexing methods but it lacks of more sophisticated approaches to calculate joins.
For SPARQLGX, almost all the queries are computed in around twenty seconds, except the Complex ones (C) that have a higher number of resulting tuples. PRoST outperforms SPARQLGX in every case, mostly by around an order of magnitude. Since SPARQLGX uses simple vertical partitioning, this difference confirms further the improvements produced by the introduction of the new data organization approach and also increases the validity of the implementation choices of PRoST.
The evaluation results presented in \cite{graux2016sparqlgx} are obtained using a virtual cluster of 10 nodes on 2 physical machines. 
In particular, when SPARQLGX is compared to S2RDF, their results are not consistent with the results from our experiment, obtained on a cluster of 10 physical machines.

    \section{Conclusion and Future Work}
\label{sec:conclusionfuturework}

In this paper, we presented PRoST, a distributed system for RDF storage and SPARQL querying built on top of Apache Spark. With PRoST we introduce an innovative data structure for partitioning RDF data that, to the best of our knowledge, for the first time combines efficiently two pre-existing approaches, namely Vertical Partitioning and the Property Table.
Our evaluations, in which we compare PRoST on a Hadoop cluster with some state-of-the-art systems Rya, SPARQLGX, and S2RDF, show that PRoST outperforms Rya and SPARQLGX in terms of querying time to a large extent and that it achieves similar results to S2RDF.
Notably, PRoST shows consistently good results for every type of query and is not limited to datasets with certain characteristics to keep the pre-computation feasible. Therefore, our approach is in particular appropriate for 
real-world applications, 
for which 
the query type and the dataset are unknown a priori.

For future work, we considered further improvements in terms of both storage and querying performance. 
For example, a promising step might be to add another Property Table where, instead of the subjects, the rows would be created around objects. This could be beneficial for triple patterns that share the same object. Another possible improvement would be to collect more precise statistics of the input dataset in order to produce better trees and, hence, a less expensive retrieval.


\bibliographystyle{ACM-Reference-Format}
\bibliography{bib/topic1}

\end{document}